\documentclass[11pt]{article}
\usepackage{moriondnophoto,epsfig}

\def\Journal#1#2#3#4{{#1} {\bf #2}, #3 (#4)}

\def\NIMA{{\em Nucl. Instrum. Methods} A}
\def\NPB{{\em Nucl. Phys.} B}
\def\PLB{{\em Phys. Lett.}  B}

\def\SNP{\em Sov. J. Nucl. Phys.}
\def\PAN{\em Phys. At. Nucl.}
\def\Sak{\em Zh. Eksp. Theor. Fiz.}
\def\DIN{\em DIRAC Internal Note }
\def\JPB{\em J. Phys. B}

\def\be{\begin{equation}}
\def\ee{\end{equation}}
\def\bea{\begin{eqnarray}}
\def\eea{\end{eqnarray}}

\def\pionium{$A_{2\pi}$}
\begin{document}
\vspace*{4cm}
\title{THE DIRAC EXPERIMENT AT CERN}

\author{ CH. P. SCHUETZ }

\address{Institut f\" ur Physik der Universit\" at Basel,\\
Klingelbergstrasse 82, CH-4056 Basel, Switzerland\\
FOR THE DIRAC COLLABORATION}

\maketitle
\abstracts{The DIRAC experiment at CERN aims to measure the lifetime of the pionium atom ($A_{2\pi}$), a $\pi^+\pi^-$ bound state with an accuracy of 10\%. The experimental setup consists of a high precision magnetic double arm spectrometer, located at the high intensity proton beam of the CERN Proton Synchrotron. This measurement will provide - in a model independent way - the S-wave pion scattering length difference $|a_0 - a_2|$ with 5\% precision.}

\section{Introduction}
The pionium atom \pionium\ is a hydrogen-like bound state of a $\pi^+$ and a $\pi^-$ meson. This atom decays predominantely strongly into two uncharged $\pi^0\pi^0$. The transition matrix of this decay is directly proportional to the difference of the two S-wave $\pi\pi$ scattering lengths with isospin 0 and 2: $a_0-a_2$. The pionium lifetime $\tau$ is inversely proportional to the squared scattering length difference. In the framework of chiral perturbation theory (ChPT), the scattering length difference has been precisely calculated~\cite{Colangelo}.

The DIRAC experiment aims to determine the lifetime of pionium and hence the difference in the two scattering lengths to provide the possibility to check the ChPT predictions in a model independent way. The expected accuracy of the lifetime is 10\% which translates into 5\% on the scattering length difference.

\section{Lifetime measurement}
The \pionium\ atoms are produced in the proton - nucleus interaction from secondary $\pi^+$ and $\pi^-$ by Coulomb interaction in the final state~\cite{Nem,Afa}. After production the pionium atoms travel through the target material where they are exposed to Coulomb interaction with target atoms. Due to this interaction, some of the atoms break up into two free pions, a $\pi^+\pi^-$ pair (atomic pairs) with small relative c.m. momenta of $Q<3 $MeV/c. The amount of broken up atoms, $n_A$, depends on the lifetime of the atoms and on the Coulomb interaction within the target material. Therefore the breakup probability $P_{br}$ -- the number of broken up atoms $n_A$ divided by the number of produced atoms $N_A$ -- is a function of the pionium lifetime and the target material. Its dependence on $\tau$ is determined using precisely calculated excitation cross sections~\cite{trautmann} and solving differential transport equations~\cite{MCatoms,trautmann2}.

In addition, the proton - target interaction produces $\pi^+\pi^-$ pairs with (Coulomb pairs) and without (uncorrelated pairs) Coulomb final state interaction. The category 'uncorrelated pairs' includes pion pairs with one pion from the decay of long-lived resonances (non-Coulomb pairs) as well as two pions from different interactions (accidental pairs). The relative momentum distributions of the Coulomb pairs and accidental pairs are linked together by the Sakharov factor~\cite{Sak}. The number of produced atoms is related to the number of produced Coulomb pairs~\cite{Nem,kfac}.

DIRAC wants to obtain the lifetime of \pionium\ by determining $P_{br}=\frac{n_A}{N_A}$. It measures $n_A$ directly by looking for excess events at very low $Q$ and it estimates $N_A$ from the number of measured Coulomb pairs in the same low $Q$ region.

\section{Experimental setup}
The DIRAC setup~\cite{setup} consists of a double arm spectrometer optimized to detect $\pi^+\pi^-$ pairs with small relative momenta. It uses the CERN T8 proton beam (24 GeV/c). The proton beam hits the target (typically 100 $\mu$m thick foils) and continues to the beam dump. The spectrometer is inclined upwards by $5.7^\circ$ with respect to the primary beam axis and has an acceptance of $\pm 1.0^\circ$ horizontally and vertically with respect to the target. The incoming $\pi^+\pi^-$ pairs travel then in vacuum through the upstream part of the setup, which hosts a set of coordinate detectors, before they are split by a bending magnet into a positive and a negative arm. Both arms are equipped with high precision drift chambers, time of flight detectors, preshower and cherenkov and muon rejection counters. The relative timing resolution of the two arms is around 180ps. 

The momentum reconstruction makes use of the drift chamber information of the two arms as well as the measured hits of a set of coordinate detectors in the upstream part. The setup has been calibrated using the $p\pi^-$ invariant mass coming from $\Lambda$ decays. The width of the peak is 0.43 MeV/c. The relative momentum resolution was estimated using Monte Carlo data~\cite{Note0201}. The resolution of $Q_x$ and $Q_y$ are estimated to be around 0.4 MeV/c. The resolution in the longitudinal component is around $\sigma_{Q_l}= $ 0.6 MeV/c. The resolution of the total $Q$ is hence better than 1 MeV/c. A system of fast trigger processors~\cite{trigger} selects low $Q$ events.

\section{Analysis}
\subsection{Data selection}
The signal extraction of atomic pairs exploits the specific signature of such events, namely time-correlation and very small relative momentum. After rejection of electrons and muons, time correlated events are selected. In addition, only events with $|Q_x|,|Q_y| < 4$ MeV/c and $|Q_l| <$ 15 MeV/c are chosen for further processing. This reduced sample consists of atomic, Coulomb and uncorrelated pairs only. The data here presented was collected with a nickel target mainly in 2001. It represents about 40\% of our total statistics.

\subsection{Background determination}
The background determination has been performed using high statistics Monte Carlo data. The Monte Carlo data is obtained using a GEANT simulation of the setup which reproduces the same data structures as the real experiment. Specifically, Coulomb and uncorrelated pairs are generated according to their phase space using momentum distributions as measured by our spectrometer.

\subsection{Signal extraction}
\begin{figure}[!htb]
\begin{tabular}{cc}
\epsfig{figure=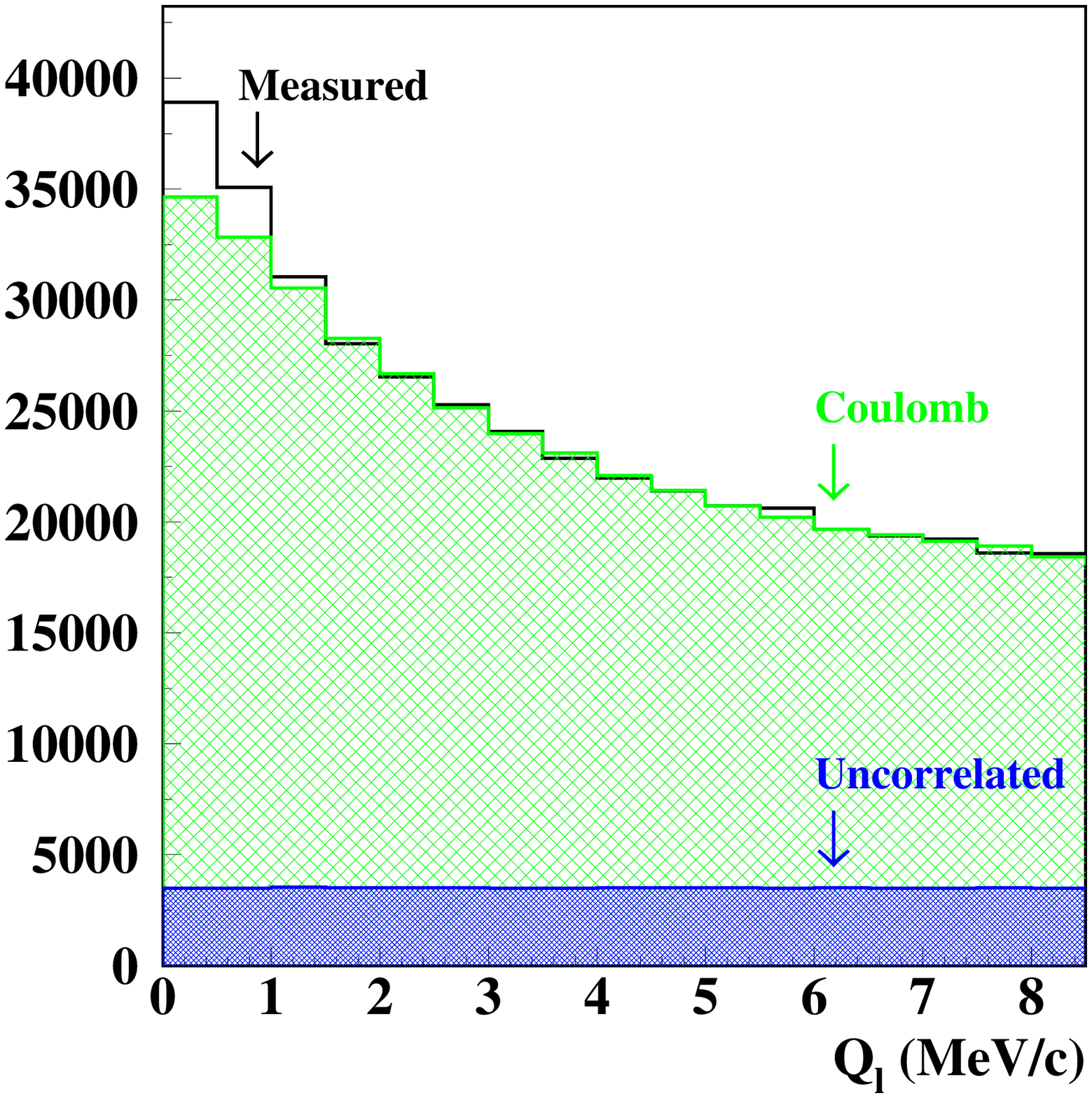,width=0.5\linewidth} &
\epsfig{figure=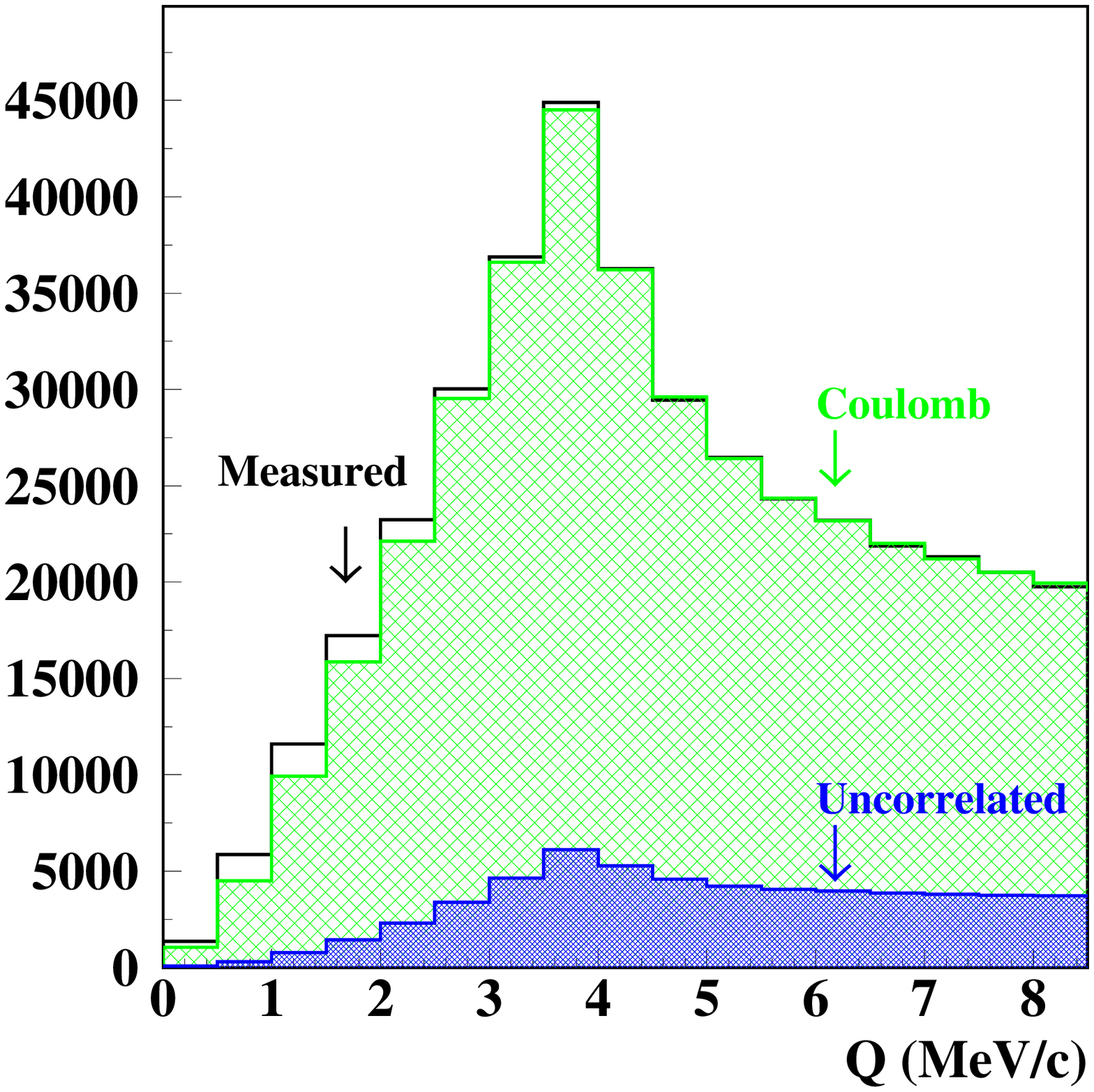,width=0.5\linewidth} \\
\end{tabular}
\caption{Measured prompt $Q_l$ and $Q_{tot}$ distributions (black) and fitted Coulomb (green) and uncorrelated (blue) background.}
\label{fig:bck}
\end{figure}

All atomic pairs are found below $|Q_l| < 2$ MeV/c and $Q_{tot} < $ 4 MeV/c. Hence, they can be extracted from the measured prompt spectrum by subtracting the background of Coulomb and uncorrelated pairs from the measured spectrum. The background contributions are determined by fitting the high statistics background to the measured spectrum in $Q_l$ and $Q_{tot}$ simultaneously above the signal region $|Q_l| > 2$ MeV/c, $Q_{tot} > $ 4 MeV/c. Figure \ref{fig:bck} displays $Q_l$ and $Q_{tot}$ distributions for measured events (black), together with the fitted uncorrelated (blue) and Coulomb (green) background contributions. The signal excess is clearly visible.
Due to the very different shapes of the Coulomb and uncorrelated distributions in $Q_l$, the relative contributions can be obtained in a reliable manner.

\begin{figure}[!htb]
\begin{tabular}{cc}
\epsfig{figure=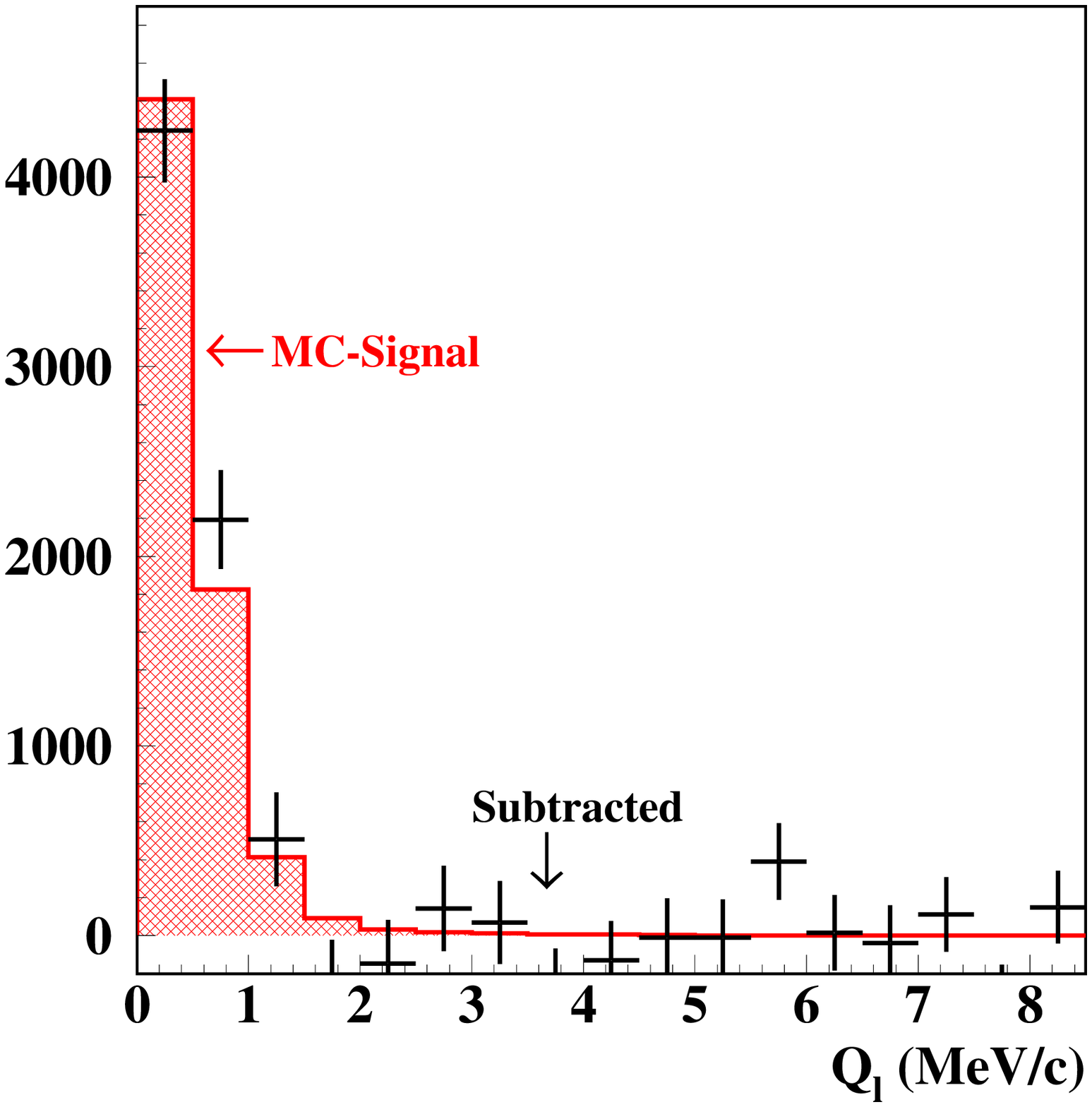,width=0.5\linewidth} &
\epsfig{figure=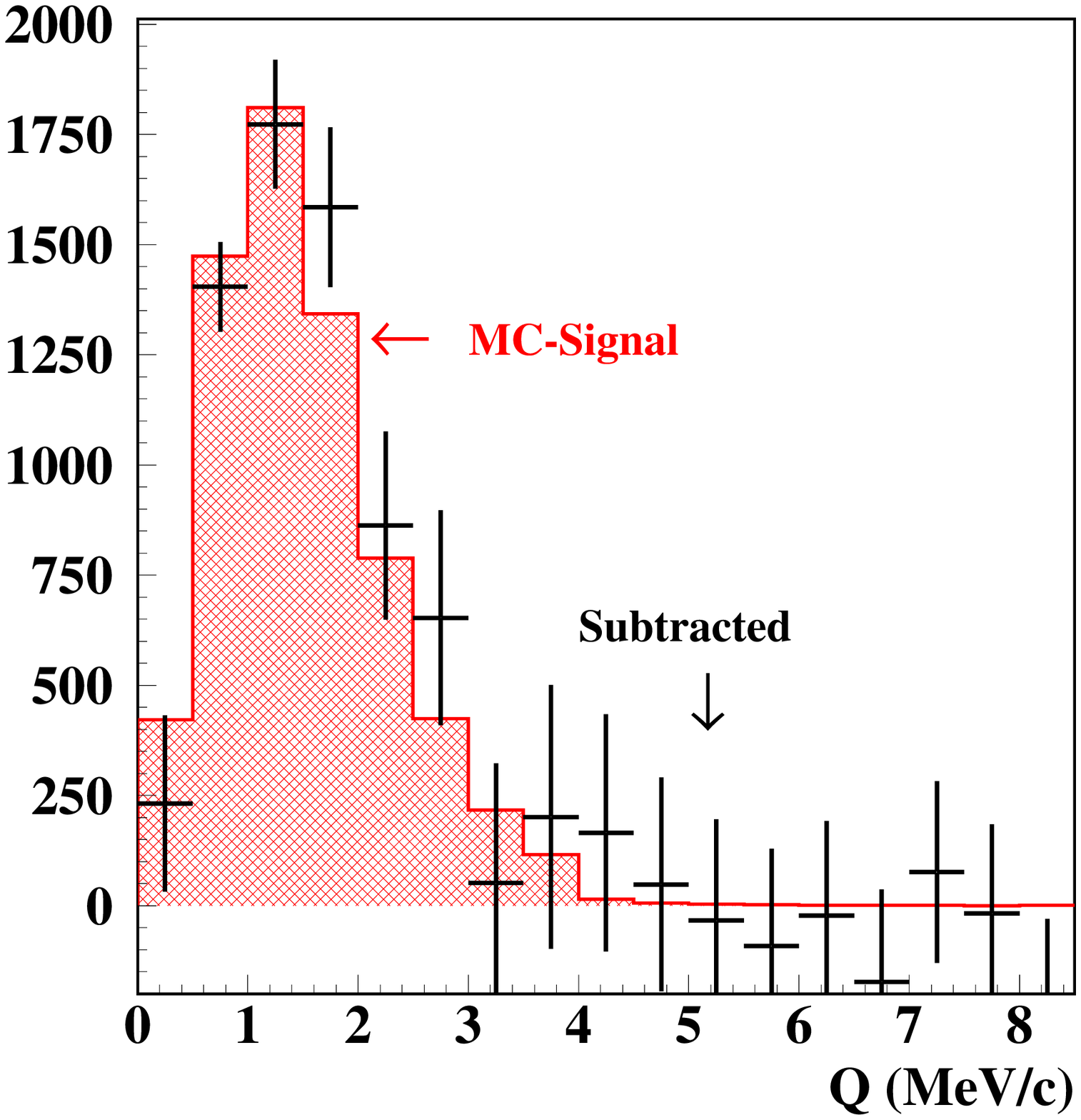,width=0.5\linewidth} \\
\end{tabular}
\caption{Residuals after background subtraction (black). For comparison is shown the simulated signal for 6800 events (red) from Monte Carlo data.}
\label{fig:signal}
\end{figure}

Subtracting the background from the measured spectrum yields the pure atomic pair signal (black) as shown in figure \ref{fig:signal}. It amounts to 6800$\pm $ 400 (stat) detected atomic pairs for the analyzed data. Since the data analysis procedure is not completed yet, this is only a preliminary result.
For comparison the expected atomic pair signal from Monte Carlo simulation is shown in red. The signal obtained from the subtraction is in good agreement with the simulated signal shape.

\subsection{Lifetime prediction}
To obtain the lifetime, the breakup probability has to be calculated, which in turn implies the knowledge of the number of broken up atoms (as measured above) and the number of produced atoms (which is related to the Coulomb background). The relation between the Coulomb background and the number of produced atoms can be calculated precisely at production. After multiple scattering (MS) in the target, however, it depends strongly on the exact multiple scattering description. To date, the best MS description for our energy range has an accuracy of 5\%. To deduce the \pionium\ lifetime reliably, DIRAC will perform dedicated multiple scattering and background measurements in 2003.

\section{Conclusion}
The DIRAC spectrometer performs well and is collecting data. Its relative momentum resolution is better than 1 MeV/c, which allows to detect $\pi^+\pi^-$ pairs from \pionium\ ionization. For the analyzed data, a background simulation using high statistics Monte Carlo data has been performed. The extracted signal amounts to 6800$\pm$ 400 (stat) atomic pairs. Due to the strong dependence on multiple scattering for the lifetime determination, DIRAC will perform dedicated measurements in 2003.

\end{document}